\documentclass[article,preprint,groupedaddress]{revtex4}
\usepackage{epsfig}
\usepackage{graphicx}
\usepackage{amssymb}

\begin{document}

\title{Universal lower bound on orbital periods around central compact objects}
%\title{Quantum-gravity bound on orbital periods around central compact objects}
\author{Shahar Hod}
\address{The Ruppin Academic Center, Emeq Hefer 40250, Israel}
\address{}
\address{The Hadassah Institute, Jerusalem 91010, Israel}
\date{\today}

\begin{abstract}

\ \ \ It is proved, using the curved line element of a spherically symmetric charged object in general relativity and 
the Schwinger discharge mechanism of quantum field theory, that the orbital periods $T_{\infty}$ 
of test particles around central compact objects as measured by flat-space asymptotic observers 
are fundamentally bounded from below. 
The lower bound on orbital periods 
becomes universal (independent of the mass $M$ of the central compact object) 
in the dimensionless $ME_{\text{c}}\gg1$ regime, 
in which case it can be expressed in terms of the electric charge $e$ and the proper mass $m_{e}$ 
of the lightest charged particle in nature: $T_{\infty}>{{2\pi e\hbar}\over{\sqrt{G}c^2 m^2_{e}}}$ (here 
$E_{\text{c}}=m^2_{e}/e\hbar$ is the critical electric field for pair production). 
The explicit dependence of the bound on the fundamental constants of nature $\{G,c,\hbar\}$ suggests 
that it may reflect a fundamental physical property of the elusive quantum theory of gravity. 
\end{abstract}
\bigskip
\maketitle

\section{Introduction}

The theory of quantum gravity is notorious for its elusiveness. 
In particular, despite the fact that the physical laws of general relativity and quantum field theory are 
well established, it is still very difficult to reveal fundamental physical principles that are expected to remain 
valid within the framework of the yet unknown quantum theory of gravity.
One such principle is the holographic entropy-area bound, whose compact 
formula $S/A\leq {{k_{\text{B}}c^3}\over{4G\hbar}}$ contains the fundamental 
constants of gravity ($G$), relativity ($c$), and quantum theory ($\hbar$) \cite{Hol1,Hol2,Hol3}.

The main goal of the present compact paper is to reveal the existence of another fundamental physical 
bound (which, admittedly, is probably far less important than the holographic entropy-area bound) 
whose formula contains the three basic constants of nature. 
In particular, below we shall explicitly prove that, in curved spacetimes, the orbital periods 
of test particles around central compact objects are bounded from below by a fundamental limit which is 
expressed in terms of the basic constants of nature: $G, c$, and $\hbar$. 

Closed circular motions of test particles around central compact objects provide valuable information 
on the non-trivial geometries of the corresponding curved spacetimes 
(see \cite{Bar,Chan,Shap} and references therein). 
In particular, an empirically measured quantity which is important for the analysis of closed 
circular motions in curved spacetimes is the orbital period $T_{\infty}$ around the central compact object 
as measured by far away asymptotic observers. 

Using a naive flat-space argument, which ignores the intriguing time-dilation/contraction 
effect of general relativity (below we shall analyze in detail the influence of this physically important 
effect on asymptotically measured orbital periods), 
it is quite easy to prove that the orbital period of a test particle around a (possibly charged) central 
compact object of radius $R$ should be bounded from below. 
In particular, since the radius of a spherically symmetric object of mass $M$ and electric charge $Q$ 
that respects the weak (positive) energy condition \cite{HawEl} is expected to be bounded from below by 
its Schwarzschild radius \cite{Notehun} ($R\gtrsim M$) and also by its classical charge 
radius ($R\geq {{Q^2}/{2M}}$) \cite{Got}, 
the orbital period $T\geq 2\pi R$ \cite{Notecc} around the central compact object as measured by inertial observers 
is expected to be bounded from below by the simple functional relation \cite{Hodfast}
\begin{equation}\label{Eq1}
T\geq T^{\text{min}}(M,Q)=2\pi\cdot{\text{max}}\{M,Q^2/2M\}\  .
\end{equation}

Intriguingly, however, it is well established \cite{Vio1,Vio2,Vio3,Vio4,Vio5,For} that, 
due to quantum coherence effects, local energy densities in {\it quantum} field theory can be negative. 
Likewise, if the matter fields inside a compact object are characterized by 
a {\it non}-minimal coupling to gravity then negative energy densities 
are not excluded even at the classical level \cite{BekMay}. 
The appearance of regions with negative energy densities inside a compact object may allow its radius $R$ 
to be {\it smaller} than its classical radius, thus opening the possibility for the existence of closed 
circular trajectories around the compact object that violate the purely classical lower bound (\ref{Eq1}).

Based on this expectation, we here raise the following physically interesting question:
Is there a {\it fundamental} quantum-gravity lower bound on orbital periods of test particles around central 
compact objects? 

The main goal of the present compact paper is to reveal the existence of such a bound 
on the orbital periods, as measured by asymptotic observers, around spherically symmetric central compact objects, 
a bound which is valid even for compact objects that may violate the classical lower bound (\ref{Eq1}). 
In particular, below we shall explicitly prove that 
the Schwinger pair-production mechanism \cite{Sch1,Sch2,Sch3}, 
a purely quantum effect, sets a lower bound on the orbital periods of test particles around 
central compact objects in the composed Einstein-Maxwell field theory. 

\section{Lower bound on orbital periods around central compact objects}

\subsection{Circular trajectories around compact objects that violate the weak energy condition}

In the present section we shall determine the shortest possible orbital period, $T^{\text{min}}_{\infty}(M)$, 
around a central compact object of total gravitational mass $M$ in the composed Einstein-Maxwell field theory. 

The external spacetime of a spherically symmetric charged compact object of radius $R$, 
total mass $M$ \cite{NoteMh}, and electric charge $Q$ \cite{NoteQ0} is characterized by the Reissner-Nordstr\"om curved 
line element \cite{Chan}
%\begin{eqnarray}\label{Eq2}
%ds^2=-\Big[1-{{2M(r)}\over{r}}\Big]dt^2+\Big[1-{{2M(r)}\over{r}}\Big]^{-1}dr^2
%+r^2d\theta^2+r^2\sin^2\theta d\phi^2\ \ \ \ \ \text{for}\ \ \ \ \ r\geq R\  ,
%\end{eqnarray}
\begin{eqnarray}\label{Eq2}
ds^2&=&-\Big[1-{{2M(r)}\over{r}}\Big]dt^2+\Big[1-{{2M(r)}\over{r}}\Big]^{-1}dr^2+\nonumber \\ &&
r^2d\theta^2+r^2\sin^2\theta d\phi^2\ \ \ \ \ \text{for}\ \ \ \ \ r\geq R\  ,
\end{eqnarray}
where
\begin{equation}\label{Eq3}
M(r)=M-{{Q^2}\over{2r}}\
\end{equation}
is the gravitational mass contained within a sphere of radius $r$.
%\begin{eqnarray}\label{Eq2}
%ds^2=-\Big(1-{{2M}\over{r}}+{{Q^2}\over{r^2}}\Big)dt^2+
%\Big(1-{{2M}\over{r}}+{{Q^2}\over{r^2}}\Big)^{-1}dr^2
%+r^2d\theta^2+r^2\sin^2\theta d\phi^2\ \ \ \ \ \text{for}\ \ \ \ \ r\geq R\  .
%\end{eqnarray}

Our goal is to determine the {\it shortest} possible 
orbital period $T^{\text{min}}_{\infty}(M)$ as measured by asymptotic observers around the central compact object. 
We shall therefore consider test particles that move arbitrarily close to the speed of light \cite{Notenng}, 
in which case the asymptotically measured orbital periods 
can be determined from the curved line element (\ref{Eq2}) with the properties \cite{Notethet}:
\begin{equation}\label{Eq4}
ds=dr=d\theta=0\ \ \ \ \  \text{and}\ \ \ \ \ \Delta\phi=\pm2\pi\  .
\end{equation}
Substituting the relations (\ref{Eq4}) into Eq. (\ref{Eq2}), one obtains the functional expression  
\begin{equation}\label{Eq5}
T_{\infty}(M,Q,r)={{2\pi r}\over{\sqrt{1-{{2M}\over{r}}+{{Q^2}\over{r^2}}}}}\
\end{equation}
for the orbital period 
around a central (possibly charged) compact object as measured by asymptotic observers.

As discussed above, classical charged compact objects that respect the weak (positive) 
energy condition \cite{HawEl} must be larger than their classical charge radius [see Eq. (\ref{Eq3})] \cite{Got}, 
\begin{equation}\label{Eq6}
R\geq R_{\text{c}}={{Q^2}\over{2M}}\  ,
\end{equation}
in which case one finds the dimensionless inequality 
$g_{tt}=1-{{2M}/{r}}+{{Q^2}/{r^2}}\leq1$ for external circular trajectories (with radii $r\geq R$) 
around the central compact object. 
Substituting this relation into Eq. (\ref{Eq5}), one obtains the 
simple {\it classical} lower bound \cite{Noterm}
\begin{equation}\label{Eq7}
T_{\infty}(M,Q,r)\geq 2\pi r\geq 2\pi R\geq2\pi\cdot{\text{max}}\{Q^2/2M,M\}\
\end{equation}
on orbital periods around central compact objects. 

However, as emphasized above, negative energy densities are not always excluded 
in physics. In particular, they may appear due to a non-minimal direct coupling of matter fields 
to gravity \cite{BekMay} and also due to quantum coherence effects in quantum field theories \cite{Vio1,Vio2,Vio3,Vio4,Vio5,For}. 
The possible existence of spacetime regions with negative energy densities inside a compact object 
that violates the (classical) weak energy condition may allow its radius $R$ to be smaller than 
its classical charge radius $R_{\text{c}}$, thus opening the possibility for the existence 
of circular trajectories, whose radii lie in the regime 
\begin{equation}\label{Eq8}
r\in[R,R_{\text{c}})\  ,
\end{equation}
that violate the classical lower bound (\ref{Eq7}) for {\it two} reasons: 
\newline
(1) The numerator of (\ref{Eq5}) is smaller than $2\pi R_{\text{c}}$ for 
circular trajectories in the regime $R\leq r<R_{\text{c}}$. 
\newline
(2) The local mass $M(R)=M-Q^2/2R$ contained within a charged compact object 
which is smaller than its classical radius is negative. 
This fact implies that the denominator of (\ref{Eq5}) is larger than $1$ for circular trajectories 
in the regime $R\leq r<R_{\text{c}}$. 
Thus, the orbital periods that characterize circular trajectories in the regime (\ref{Eq8}) are {\it blueshifted} 
by the factor $\sqrt{1-2M/r+Q^2/r^2}>1$ as measured by asymptotic 
observers. This is a general relativistic time contraction effect. 

Before proceeding, it is important to emphasize that, despite the fact that the 
local mass (\ref{Eq3}) contained within a compact object which is smaller than 
its classical radius is negative, we shall assume 
that the {\it total} ADM mass $M$ of the spacetime as measured by asymptotic observers is positive. 

\subsection{A fundamental lower bound on orbital periods around central compact objects}

It is of physical interest to explore the physical and mathematical properties of closed circular motions 
around central compact objects that, due to quantum coherence effects \cite{Vio1,Vio2,Vio3,Vio4,Vio5,For} 
or non-minimal coupling to gravity \cite{BekMay}, may violate the classical lower bound (\ref{Eq7}). 
In particular, one naturally wonders whether orbital periods around central compact objects 
that violate the weak energy condition are 
fundamentally bounded from below by the physical laws of general relativity and quantum theory? 

In the present section we shall reveal the physically intriguing fact that the 
Schwinger pair-production mechanism \cite{Sch1,Sch2,Sch3} 
sets a fundamental quantum lower bound on the orbital period (\ref{Eq5}) around central compact objects, 
a bound which is valid even for compact objects that violate the weak energy condition and can therefore 
violate the classical lower bound (\ref{Eq7}).% \cite{Notenff}. 

In particular, we shall now prove that, as opposed to the unbounded 
redshift (time dilation) effect which characterizes the orbital periods of test particles that circle a central black hole close 
to its horizon [$T_{\infty}(r\to R_{\text{horizon}})\to\infty$], 
the blueshift time contraction effect, which characterizes the orbital 
periods of test particles that circle a central compact object with negative energy densities \cite{Vio1,Vio2,Vio3,Vio4,Vio5,For,BekMay}, is fundamentally (quantum mechanically) bounded 
from above. 

Our goal is to determine the shortest possible orbital period $T^{\text{min}}_{\infty}(M)$ 
around a central compact object of a given mass $M$. 
To this end, we first point out that, for given values of the gravitational mass $M>0$ of 
the central compact object and the radius $r$ of the external circular trajectory, 
the orbital period $T_{\infty}(Q;M,r)$ as measured by asymptotic 
observers decreases monotonically with the electric charge $Q$ of the central compact object [see 
Eq. (\ref{Eq5})]. 
Thus, in order to minimize the orbital period (\ref{Eq5}) of a test particle 
around a central object with a given total mass $M$, 
one should maximize its electric charge. 
In particular,
\begin{equation}\label{Eq9}
T^{\text{min}}_{\infty}(M,Q_{\text{max}},r)=
{{2\pi r}\over{\sqrt{1-{{2M}\over{r}}+{{Q^2_{\text{max}}}\over{r^2}}}}}\  ,
\end{equation}
where $Q_{\text{max}}=Q_{\text{max}}(r)$ is the maximally allowed electric charge that can be 
contained within a sphere of radius $r$.

One naturally wonders: What physics prevents us from making the expression (\ref{Eq9}) for the orbital period 
as small as we wish? 
Or, in other words, we ask: 
Is there a fundamental physical mechanism that bounds the electric charge $Q_{\text{max}}(r)$ 
that can be contained within a sphere of radius $r$? The answer is `yes'! 

In particular, the Schwinger pair-production mechanism (a {\it quantum} polarization effect) implies 
the existence of a fundamental upper bound on the
electric field strength of the charged compact object \cite{Sch1,Sch2,Sch3,Noteoom}:
\begin{equation}\label{Eq10}
{{Q}\over{r^2}}\leq E_{\text{c}}\equiv {{m^2_e}\over{e\hbar}}\  ,
\end{equation}
where $\{e,m_e\}$ are respectively the electric charge and the proper mass 
of the lightest charged particle in nature. 
Substituting the upper bound (\ref{Eq10}) into Eq. (\ref{Eq9}), one obtains the radius-dependent 
functional expression
\begin{equation}\label{Eq11}
T^{\text{min}}_{\infty}(r;M,E_{\text{c}})={{2\pi r}\over{\sqrt{1-{{2M}\over{r}}+E^2_{\text{c}}r^2}}}\
\end{equation}
for the shortest possible orbital period.

Interestingly, and most importantly for our analysis, one finds from Eq. (\ref{Eq11}) that, 
for a given total mass $M$ of the system, 
the orbital period is {\it minimized} for the field-independent ($E_{\text{c}}$-{\it independent}) orbital radius
\begin{equation}\label{Eq12}
r^{\text{min}}=3M\  .
\end{equation}

Since the electric charge is confined to the interior of the central compact object, the electric field strength 
in the exterior ($r\geq R$) spacetime region is a monotonically decreasing function of the orbital 
radius $r$. Thus, the assumption in Eq. (\ref{Eq11}) that the electric field along the trajectory of the test particle saturates 
the quantum upper bound (\ref{Eq10}) [as discussed above, for a given value $r$ of the orbital radius, 
the larger the electric field along the trajectory of the particle, the shorter is the orbital period as measured by asymptotic observers, see Eq. (\ref{Eq11})] corresponds to 
the assumption that the particle moves along a circular trajectory 
infinitesimally close to the surface of the charged compact object: 
\begin{equation}\label{Eq13}
r^{\text{min}}\to R^+\  .
\end{equation}

Substituting (\ref{Eq12}) into Eq. (\ref{Eq11}), one obtains the remarkably simple functional relation
\begin{equation}\label{Eq14}
T^{\text{min}}_{\infty}(M,E_{\text{c}})=2\pi\cdot\sqrt{{27M^2}\over{1+27M^2E^2_{\text{c}}}}\
\end{equation}
for the shortest possible orbital period around a central compact object of total 
gravitational mass $M$ as measured by asymptotic observers. 

It is interesting to note that, in the dimensionless regime $ME_{\text{c}}\ll1$, 
the analytically derived lower bound (\ref{Eq14}) on orbital periods around central compact objects 
yields the {\it classical} ($\hbar$-independent) bound
\begin{equation}\label{Eq15}
T^{\text{min}}_{\infty}(M,E_{\text{c}})\ \to\ 6\sqrt{3}\pi\cdot M\ \ \ \ \ \text{for}\ \ \ \ \ ME_{\text{c}}\ll1\  .
\end{equation}
On the other hand, in the opposite dimensionless regime $ME_{\text{c}}\gg1$ the lower bound (\ref{Eq14}) 
yields the purely {\it quantum} ($\hbar$-dependent) bound 
\begin{equation}\label{Eq16}
T^{\text{min}}_{\infty}(M,E_{\text{c}})\ \to\ {{2\pi}\cdot{E^{-1}_{\text{c}}}}
\ \ \ \ \ \text{for}\ \ \ \ \ ME_{\text{c}}\gg1\  .
\end{equation}
Intriguingly, the lower bound (\ref{Eq16}) is universal in the sense that it is {\it independent} of the 
mass $M$ of the central compact object. 

It is worth stressing the fact that, in the dimensionless $ME_{\text{c}}\gg1$ regime, 
the value of $T^{\text{min}}$, as given by the analytically derived quantum 
expression (\ref{Eq16}), satisfies the strong inequality $T^{\text{min}}(M,E_{\text{c}})={{2\pi}/{E_{\text{c}}}}\ll 2\pi\cdot{\text{max}}\{Q^2/2M,M\}$ \cite{NoteQM}, and it therefore violates 
the classical bound (\ref{Eq7}) on orbital periods. 

\section{Summary and Discussion}

Motivated by the fact that negative energy densities may appear in various physical situations 
(for example, due to quantum coherence effects that appear in quantum 
field theories \cite{Vio1,Vio2,Vio3,Vio4,Vio5,For} and also due to a possible non-minimal coupling of matter fields 
to gravity \cite{BekMay}), in the present compact paper 
we have raised the following physically important question:
Is there a {\it fundamental} lower bound on the orbital periods of test particles 
around central compact objects? 

In order to address this question, we have analyzed the three-dimensional \cite{Notethr} functional behavior 
of the orbital periods $T_{\infty}=T_{\infty}(M,Q,r)$ of test particles whose velocities are arbitrarily close to the 
speed of light on the physical parameters $\{M,Q\}$ 
that characterize the central compact object and on the radii $r$ of the external circular trajectories. 

In particular, the main goal of the present paper was to derive a robust lower bound on orbital periods of 
test particles around spherically symmetric central compact objects, 
a bound which is valid even for compact objects that violate the weak (positive) energy condition and thus 
may violate the classical lower bound (\ref{Eq6}) on the radii of charged compact objects and 
the corresponding classical lower bound (\ref{Eq7}) on orbital periods as 
measured by asymptotic observers. 

Intriguingly, we have revealed the fact that the Schwinger pair-production mechanism \cite{Sch1,Sch2,Sch3}, 
a purely quantum effect, is responsible for the existence of a previously unknown fundamental 
lower bound on the orbital periods of test particles around central compact objects. 

The main analytical results derived in this paper and their physical implications are as follows:

(1) We have emphasized the fact that external circular trajectories around (possibly charged) central compact objects 
that respect the classical weak (positive) 
energy condition are characterized by the relation $1-{{2M}/{r}}+{{Q^2}/{r^2}}<1$ [see the 
lower bound (\ref{Eq6})], in which case the orbital periods as measured by far away asymptotic observers 
are longer than the corresponding locally measured orbital times (this 
is the familiar time dilation effect in general relativity). 

On the other hand, we have pointed out that charged compact objects that violate the classical positive energy 
condition may be characterized by the presence of external circular trajectories in the regime $r\in[R,R_{\text{c}})$ 
with the property $1-{{2M}/{r}}+{{Q^2}/{r^2}}>1$, in which case the general relativistic time contraction effect implies 
that the asymptotically measured orbital periods are {\it shorter} than the corresponding locally 
measured orbital periods.  

Interestingly, we have explicitly proved that, as opposed to the unbounded time dilation (redshift) 
effect, $T_{\infty}(r\to R_{\text{horizon}})\to\infty$, which characterizes the orbital periods of test particles 
in the near-horizon region of a central black hole, the time contraction (blueshift) effect, 
which characterizes the orbital periods of test particles around central compact objects with negative energy densities, 
is fundamentally (quantum mechanically) bounded from above according to the analytically derived functional 
relation (\ref{Eq14}). 

(2) Using the curved line element (\ref{Eq2}) that characterizes the exterior spacetime region of a 
spherically symmetric charged compact object in general relativity and 
the Schwinger discharge mechanism of quantum field theory, we have explicitly proved 
that the orbital periods $T_{\infty}(M)$ of 
test particles around a central compact object of total 
mass $M$ as measured by asymptotic observers are 
fundamentally bounded from below by the functional relation [see Eqs. (\ref{Eq10}) and (\ref{Eq14})]
\begin{equation}\label{Eq17}
T_{\infty}(M)\geq T^{\text{min}}_{\infty}(M)=
2\pi\cdot\sqrt{{27M^2}\over{1+27M^2\cdot({{m^2_e}/{e\hbar}})^2}}\  .
\end{equation}
For a central object of total mass $M$, the minimally allowed orbital period (\ref{Eq17}) is 
obtained for the following physical parameters of the compact object and the 
circular trajectory: $Q=9M^2E_{\text{c}}$ and $r\to R=3M$ 
[see Eqs. (\ref{Eq10}), (\ref{Eq12}), and (\ref{Eq13})]. 

It is worth emphasizing the fact that, as opposed to the classical lower bound (\ref{Eq7}) on orbital periods, 
the analytically derived lower bound (\ref{Eq17}) is valid even in the quantum regime of 
matter fields that may violate the weak energy condition. 
In particular, circular trajectories around central compact objects whose radii violate the 
classical lower bound (\ref{Eq6}) are still characterized by the fundamental 
quantum lower bound (\ref{Eq17}) on their orbital periods.  

(3) The lower bound (\ref{Eq17}) becomes universal ({\it independent} of the mass $M$ of the central compact object) 
in the dimensionless $ME_{\text{c}}\gg1$ regime, in which case it can be expressed in the remarkably compact form 
(we write here the explicit dependence of $T^{\text{min}}_{\infty}$ on 
the fundamental constants of nature $\{G,c,\hbar\}$):
\begin{equation}\label{Eq18}
T^{\text{min}}_{\infty}\to {{2\pi\hbar\sqrt{k}e}\over{\sqrt{G}c^2 m^2_{e}}}
\ \ \ \ \ \text{for}\ \ \ \ \ ME_{\text{c}}\gg1\  ,
\end{equation}
where $\epsilon_0=1/4\pi k$ is the electric constant (vacuum permittivity). 

The explicit dependence of the minimally allowed orbital time $T^{\text{min}}_{\infty}$ 
on the fundamental constants of gravity ($G$), relativity ($c$), and quantum physics ($\hbar$) 
suggests that it may reflect a genuine physical property of a fundamental quantum theory of gravity. 

(4) Interestingly, inspection of Eq. (\ref{Eq18}) reveals the fact that, 
in order for the smallest possible orbital period $T^{\text{min}}_{\infty}$ to be larger than the fundamental 
scale set by the Planck time $t_{\text{P}}=(\hbar G/c^5)^{1/2}$, 
one must demand the existence of a {\it weak-gravity} bound of the form 
\begin{equation}\label{Eq19}
{{Gm^2_{e}}\over{ke^2}}\lesssim \alpha^{-1/2}\  ,
\end{equation}
where $\alpha$ is the dimensionless fine-structure constant.

\

%\bigskip
\noindent
{\bf ACKNOWLEDGMENTS}
%\bigskip

This research is supported by the Carmel Science Foundation. I would
like to thank Yael Oren, Arbel M. Ongo, Ayelet B. Lata, and Alona B.
Tea for helpful discussions.


\begin{thebibliography}{99}
\bibitem{Notehun} We shall henceforth use gravitational units in which $G=c=1$.
\bibitem{Notecc} Note that the equality sign in the flat-space relation $T\geq 2\pi R$ 
corresponds to a test particle that moves around the central compact object at the speed of light.
\bibitem{NoteMh} We shall assume that the central object is macroscopic with $M\gg\hbar^{1/2}$.
\bibitem{NoteQ0} We shall henceforth assume, without loss of generality, that the electric charge 
of the central compact object is characterized by the relation $Q\geq0$.
\bibitem{Notenng} Note that the orbiting particle may use non-gravitational forces 
in order to circle the central compact object on a non-geodesic trajectory with a velocity that, in principle, 
can be made arbitrarily close to the speed of light.
\bibitem{Notethet} One may assume, without loss of generality, the polar angular 
relation $\theta=\pi/2$ for equatorial trajectories.
\bibitem{Noterm} Here we have used the fact that a charged compact object with $Q\leq M$ that respects 
the weak (positive) energy condition \cite{HawEl} is larger than its Reissner-Nordstr\"om 
radius $M+(M^2-Q^2)^{1/2}\geq M$. Charged objects that respect 
the weak energy condition are characterized by the inequalities $R\geq Q^2/2M>M/2$ in the 
complementary regime $Q>M$.
\bibitem{Noteoom} It should be noted that the expression (\ref{Eq10}) for the 
critical electric field should be regarded as an order-of-magnitude estimate.
\bibitem{NoteQM} Here we have used the fact that the bound (\ref{Eq16}) is valid in the regime 
$Q/M\gg1$ with $ME_{\text{c}}\gg1$, which implies the series of strong 
inequalities $Q^2/2M\gg Q\gg M\gg 1/E_{\text{c}}$.
\bibitem{Notethr} We use here the term `three-dimensional' in order to emphasize the fact that the asymptotically 
measured orbital periods $T_{\infty}(M,Q,r)$ depend on three physical parameters: the mass $M$ of the 
central compact object, its electric charge $Q$, and the radius $r$ of the external circular orbit.
\end{thebibliography}
\end{document}